\RequirePackage{lineno}
\documentclass[twocolumn]{revtex4}

\usepackage{amsmath,amsfonts,amssymb}
\usepackage{graphicx}
\usepackage{natbib}
\setcitestyle{authoryear,round}

\usepackage[sectionbib]{bibunits}
\defaultbibliographystyle{ecol_let}

\usepackage{lineno}  

\defaultbibliography{aa_macrobib}

\begin{document}

\title{Bioenergetic trophic trade-offs determine mass-dependent extinction thresholds across the Cenozoic}

\author{
Justin D. Yeakel$^{1,2,*}$, Matthew C. Hutchinson$^{1}$, Christopher P. Kempes$^{2}$, Paul L. Koch$^{3}$, Jacquelyn L. Gill$^{4}$, Mathias M. Pires$^{5}$}

\address{
$^{1}$Life \& Environmental Sciences, University of California Merced, Merced, California, USA\\
$^{2}$The Santa Fe Institute, Santa Fe, New Mexico, USA\\
$^{3}$Earth \& Planetary Sciences, University of California Santa Cruz, Santa Cruz, California, USA\\
$^{4}$University of Maine, Derry, Maine, USA\\
$^{5}$Universidade Estadual de Campinas, Campinas, Brazil\\
$^{*}$Corresponding author: jyeakel@ucmerced.edu}







\begin{abstract}
Body size drives the energetic demands of organisms, constraining trophic interactions between species and playing a significant role in shaping the feasibility of species' populations in a community. On macroevolutionary timescales, these demands feed back to shape the selective landscape driving the evolution of body size and diet. We develop a theoretical framework for a three-level trophic food chain -- typical for terrestrial mammalian ecosystems -- premised on bioenergetic trade-offs to explore mammalian population dynamics. Our results show that interactions between predators, prey, and external subsidies generate instabilities linked to body size extrema, corresponding to observed limits of predator size and diet. These instabilities generate size-dependent constraints on coexistence and highlight a feasibility range for carnivore size between 40 to 110 kg, encompassing the mean body size of terrestrial Cenozoic hypercarnivores. Finally, we show that predator dietary generalization confers a selective advantage to larger carnivores, which then declines at megapredator body sizes, aligning with diet breadth estimates for contemporary and Pleistocene species. Our framework underscores the importance of understanding macroevolutionary constraints through the lens of ecological pressures, where the selective forces shaping and reshaping the dynamics of communities can be explored.
\end{abstract}

\maketitle

\section{Introduction}

\begin{bibunit}

Terrestrial mammalian carnivores span a wide range of body sizes, from smaller-bodied mustelids with generalist diets, including vertebrates, invertebrates, and plants, to larger-bodied vertebrate specialists such as wolves and lions.  
Body size drives the energetic demands of organisms, governing changes in resource assimilation \citep{hou2008,bhat2020scaling} and energetic allocation to somatic growth, maintenance, and reproduction \citep{West:1997cg,Kempes:2012hy} across individual lifetimes \citep{Lindstedt:2002td}. 
Whereas organismal metabolic rates are central to determining which resources are energetically rewarding and how much an organism must eat to meet its energetic demand \citep{Carbone:1999ju,hou2008}, body size also imposes biomechanical constraints on the size of prey that can be subjugated and consumed \citep{sorkin2008biomechanical}. 
The viability of a predator population is thus determined by the balance between a species' energetic demands and the ability of its prey to sustain predation while avoiding collapse \citep{otto2007allometric,Carbone:2007dz}.  
Larger predators have high energetic demands and benefit from exploiting larger herbivores that serve as prey to fewer potential competitors \citep{ritwika2024beyond}. 
Because larger mammals have slower rates of gestation and reproduction, and consequentially lower population densities, these prey species may be particularly vulnerable to increased predation rates \citep{Brook2005,rallings2024}. 

While the centrality of body size in structuring ecological systems is clear, how species interactions may have contributed to size evolution throughout the Cenozoic is less so.
Long after the rapid expansion of mammalian body size classes following the K-Pg extinction event 66 million years ago \citep{Alroy:1998p1594,Smith:2011cza,clauset2008evolution}, the cooler, dryer, and increasingly open habitat of a post-Eocene world \citep{Stromberg2011,Andermann2022} coincided with the appearance of maximal body sizes among herbivorous (at ca. 17400 kg) and carnivorous (at ca. 1000 kg) mammalian lineages across multiple continents \citep{Smith:2010p3442}. 
With the evolution of these increasingly large mammalian species came the associated bioenergetic boundaries governing the feasibility of particular interactions \citep{weitz2006size}.
These bioenergetic boundaries may be explicitly connected to population instabilities, where the requirements of a predator species may suppress its prey to the point of collapse -- a function of the body sizes and associated population densities of each \citep{yeakel2018dynamics,rallings2024}.
A theoretical framework that unites the bioenergetic constraints emerging from mammalian energetic trade-offs with the dynamics of interacting species may provide important insight into the selective drivers shaping mammalian body size distributions -- both past and present. 


To understand how the bioenergetic trade-offs associated with growth and mortality among predators and prey may impact the feasibility of species interactions and species coexistence, we construct a three-level trophic (e.g., tri-trophic) dynamic model premised on the scaling of energetic demands with body size.
By incorporating fundamental energetic trade-offs in the vital rates governing animal growth and mortality, we aim to reconstruct the ecological drivers influencing the evolution of body size.
By analyzing a tri-trophic food chain - first with one herbivore prey and one predator that is also assimilating energy from both the herbivore and a non-prey subsidy and then one with two competing herbivores and a predator, our framework identifies energetic boundaries that determine the likelihood of population collapse as potential drivers of size selection.
We then leverage our theoretical model to assess whether and to what extent body size impacts the potential for coexistence, as well as the constraints these size-based dynamics introduce to the dietary flexibility of the predator.
We show that these dynamics likely influenced how species at different trophic levels faced distinct risks, shaping the structure and function of Cenozoic mammalian communities.

\section*{Materials and Methods}

Our primary aim is to model the flow and storage of biomass between populations of interacting mammalian species as a function of body size and examine whether and to what extent energetic limitations constrain the feasibility of interacting populations.
Using mechanistic models of metabolic trade-offs governing mammalian ontogentic growth, reproduction, and mortality, combined with allometric relationships capturing body size constraints on physiological and ecological traits, we describe a flexible tri-trophic framework premised on the body sizes of each species.
We then describe how the interactions between predators and herbivore prey are also body size-dependent, and how empirical observations are used to condition interactions between the predator and its primary prey.
We present detailed allometric derivations of the functions describing mammalian vital rates in Appendix 1.
See the archived Zenodo repository \citep{Yeakel2024} to access the supplementary electronic data and code.

\subsection*{Allometric tri-trophic food chains}

Here we describe a tri-trophic food chain, consisting of a primary producer, one or more herbivore prey, and a predator.
The energy available for somatic growth, maintenance, and reproduction of the predator population $P$ is limited by its consumption of $n$ herbivore prey populations ${H_i}$, where $n$ denotes the number of prey species available to the predator. 
Each prey population is supported by the consumption of a plant resource $R$. 
The growth of the predator population is fueled by the mortality it inflicts on its prey with rate $w_i\lambda_P/Y_P$, where $w_i$ is the proportional reliance of the predator on herbivore $i$, $\lambda_P$ is the predator's effective growth rate, and $Y_P$ is the predator yield coefficient, or the grams of predator produced per gram of prey consumed.
We assume the predator's effective growth rate follows a linear (Type I) functional response across the herbivore population density ${H_i}$, and is maximized at $\lambda_{H_i}^{\rm max}$.
When the predator does not rely completely on its herbivore prey (i.e., $\sum_i w_i < 1$), the remainder of predator growth $w_S$ is obtained from a constant subsidy with availability $S$, which represents any other alternative prey or resource. 
The parameter $w$, as applied to subsidies ($w_S$) or herbivore prey ($w_i$), thus describes dietary generalization across these potential foods, where a value $w\approx1$ implies specialization on a specified resource, whereas $w<1$ implies increased generalization across multiple resources.
Below we will consider two scenarios with this general framework: 
\emph{i}) the subsidy model, where the predator consumes a single herbivore prey $(n=1)$ and is otherwise supported by an external subsidy $(S>0)$;
\emph{ii}) the competition model, where the predator is supported by two herbivore prey species $(n=2)$ without the aid of an external subsidy ($S=0$).

The rate of consumption fueling the herbivore prey population(s) is proportional to the resource density and is given by $\lambda_{H_i}(R)/Y_{H_i}$, where $\lambda_{H_i}(R)$ is the effective herbivore growth rate and $Y_{H_i}$ is the herbivore yield coefficient, or the grams of herbivore produced per gram of resource consumed \citep{Kempes:2012hy}.
We assume the herbivore's growth rate $\lambda_{H_i}(R)$ follows a saturating (Type II) functional response across the resource density $R$, where the maximum growth is $\lambda^{\rm max}_{H_i}$ and the resource half-saturation density is $\hat{k} = k/2$ \citep{rallings2024}, where $k$ is the resource carrying capacity. 
The resource grows logistically with growth rate $\alpha$ towards its carrying capacity, both of which are assumed to be characteristic of tropical grasses \citep[see Table 1;][]{yeakel2018dynamics}.

Predator mortality is the product of natural mortality $\mu_P$, which is composed of both initial cohort mortality and the cumulative effects of senescence \citep{CalderIII:1983jd,rallings2024}.
Herbivore mortality includes the effects of natural mortality, $\mu_{H_i}$, as well as starvation mortality $\sigma_{H_i}(R) = \sigma_{H_i}^{\rm max}(1- R/k)$ -- which is inversely proportional to resource density \citep{yeakel2018dynamics,rallings2024} -- and predation.
The general tri-trophic system can then written as
\begin{align}
    \frac{\rm d}{\rm dt}P &= \bigg(\sum_{i=1}^n w_i\lambda_P H_i + w_S\lambda_P^{\rm max} S\bigg)P - \mu_P P, \nonumber \\
    \frac{\rm d}{\rm dt}H_i &= \lambda_{H_i}(R)H_i - \bigg(\mu_{H_i} + \sigma_{H_i}(R) \bigg)H_i - w_i\frac{\lambda_P}{Y_P}PH_i, \nonumber \\
    \frac{{\rm d}}{{\rm dt}}R &= \alpha \left(1 - \frac{R}{k} \right)R - \sum_{i=1}^n\frac{\lambda_{H_i}(R)}{Y_{H_i}}H_i,
    \label{eq:3d}
\end{align}
where $\lambda_{H_i}(R) = \lambda_{H_i}^{\rm max}\frac{R}{\hat{k} + R}$ and $\lambda_P = \lambda_P^{\rm max}/Y_{H_i}k$.
The rate laws describing the growth and mortality of both mammalian predator and herbivore species vary as a function of predator body mass $M_P$ and herbivore body mass $M_{H}$. 
(Throughout, we will use the term $M_H$ to refer to herbivore prey body size generally, and $M_{H_i}$ to refer to prey species that are specific to the subsidy and competition models.)
We approach the derivation of vital rates with respect to predator and herbivore mass by solving for multiple timescales associated with ontogenetic growth, maintenance, and expenditure, based on the bioenergetic trade-offs associated with somatic growth and maintenance during ontogenetic growth.
See the Appendix 1 for our derivation of vital rates using this approach, also detailed in \citet{yeakel2018dynamics} and \citet{rallings2024}.

Whether or not two species engage in a trophic interaction given their respective traits drives the structures of food webs.
Trophic interactions are often constrained by body size \citep{Sinclair2003,Brose2005,Hatton:2015fk}, with large prey generally suffering mortality from large predators, though the nature of predator-prey mass relationships varies across communities \citep{barnes2010global,nakazawa2017individual} and organismal body size \citep{Brose2005,Rohr2010,riede2011stepping,Yeakel2014,pires2015pleistocene}, and may be driven by the optimization of handling time among predators \citep{delong2020detecting}.
Because the timescales of vital rates are mechanistically derived from energetic trade-offs characterizing the metabolic processes of both predators and their herbivore prey, such an approach requires that predator and herbivore masses relate in ecologically meaningful ways.
The relationship that determines how the herbivore's mass $M_H$ relates to the predator's mass $M_P$ is provided by ${\rm E}\{M_H|M_P\}$, where the function ${\rm E}\{\}$ denotes the expectation, which we obtain from observational data.
${\rm E}\{M_H|M_P\}$ follows a power law across $M_P$, where the slope 
has been shown to be roughly linear \citep{Carbone:1999ju} or slightly superlinear \citep{Uiterwaal2022}, with theoretical expectations from resource use optimization predicting a slope of ca. 1.23 \citep{delong2020detecting}.
By integrating high resolution data sets for large-bodied mammalian carnivores \citep{hayward2005lion,Hayward2006hyena,hayward2006leopard,hayward2006lycaon,hayward2006cheetah,Hayward2008}, this slope increases slightly to ca. 1.46 (see Appendix 2).

\begingroup
\renewcommand{\baselinestretch}{1} 
\begin{table}[h]
    \centering
    \footnotesize{
    \begin{tabular}{l l l}
        \hline
        \textbf{Parameter} & \textbf{Description} & \textbf{Value/Units} \\
        \hline
        \( P \) & Predator population & kg/m${}^2$ \\
        \( S \) & Constant subsidy & g/m${}^2$ \\
        \( n \) & Number of herbivore prey & \\
        \( H_1 \) & Primary herbivore prey density & kg/m${}^2$ \\
        \( H_2 \) & Secondary herbivore prey density & kg/m${}^2$ \\
        \( R \) & Plant resource & g/m${}^2$ \\
        \( w_S \) & Predator reliance on subsidy & 0:1 \\
        \( w_1 \) & Predator reliance on primary prey & 0:1 \\
        \( \lambda_P \) & Predator effective growth rate & 1/s \\
        \( Y_P \) & Predator yield coefficient & n.a. \\
        \( \mu_P \) & Predator natural mortality rate & 1/s \\
        \( M_P \) & Predator body mass & kg\\
        \( \lambda_{H_i}(R) \) & Herbivore $i$ growth rate & 1/s\\
        \( Y_{H_i} \) & Herbivore $i$ yield coefficient & n.a.\\
        \( \lambda^{\rm max}_{ {H_i}} \) & Herbivore $i$ maximum growth rate & 1/s\\
        \( \mu_{H_i} \) & Herbivore $i$ natural mortality & 1/s\\
        \( \sigma_{H_i}(R) \) & Herbivore $i$ starvation mortality rate & 1/s\\
        \( M_{H_1} \) & Primary herbivore prey body mass & \({\rm E}\{M_H|M_P\}\) kg\\
        \( M_{H_2} \) & Secondary herbivore prey body mass & \(\phi M_{H_1}\) kg\\
        \(\phi\) & Size scaling of the secondary prey & 0.1:2.0 \\
        \( \alpha \) & Resource growth rate & \(9.45\times10^{-9}~{\rm s}^{-1}\)\\
        \( k \) & Resource carrying capacity & \(2.3\times10^4~{\rm g~m}^{-2}\)\\
        \hline
    \end{tabular}
    }
    \caption{List of parameters and their meanings (Eq. \ref{eq:3d}). For derivations of mass-specific functions, see Appendix 1.}
    \label{tab:parameters}
\end{table}
\endgroup

We examine two particular model variations of Equation \ref{eq:3d}: \emph{i}) where there is a single resource, herbivore prey, and predator species that has access to an external subsidy (the subsidy model, where $n=1$ and $S>0$), and \emph{ii}) where there is a single resource, two competing herbivore prey, and a predator species without external subsidization (the competition model, where $n=2$ and $S=0$).
For the latter, we assume that the primary herbivore prey has a body mass that is determined from the relationship given by $M_{H_1} = {\rm E}\{M_{H_1}|M_P\}$.
In contrast, we assume the secondary herbivore prey has a body mass given by $M_{H_2} = \phi M_{H_1}$, and where changing the value of $\phi$ allows the secondary prey to range from smaller to larger than the primary prey.
Throughout, we use the prefix `mega' to refer to species with body sizes $\geq 600$ kg \citep{hayward2005lion}.

\section*{Results}

\subsection*{Subsidized food chains and population thresholds}

We observe that the subsidy model (Eq. \ref{eq:3d}), where the number of herbivore prey $n=1$, subsidy availability $S>0$, and where the proportion of subsidies $w_S$ and the single herbivore prey $w_1$ to the diet of the predator sum to unity, has 7 fixed points, but only one stable internal steady state for each particular combination of predator and herbivore body masses (Fig. \ref{fig:thresholdsub}).
This means that there is only one feasible steady state where species' populations coexist, such that the population densities as well as the diversity of the system depend on the assumed body sizes of the herbivore and predator.
Moreover, the internal steady state exists both in the absence and presence of the external subsidy.
Despite the appearance of instabilities occurring at specific predator and herbivore body sizes (see below), the stable steady state densities of both fall within the range of empirical mammalian densities (Fig. \ref{fig:thresholdsub}A,B), in line with similar investigations of mammalian dynamics that include mechanistic models of energetic trade-offs \citep{yeakel2018dynamics,bhat2020scaling,rallings2024}.
We note that while the steady state solutions for the predator ($P^*$) and herbivore prey ($H^*$) can be analytically expressed, they cannot be written in a compact form, and we do not report them here (see Supplemental Electronic Data in \citet{Yeakel2024}).

\begingroup
\renewcommand{\baselinestretch}{1} 
\begin{figure*}[ht]
    \centering
    \includegraphics[width=0.8\linewidth]{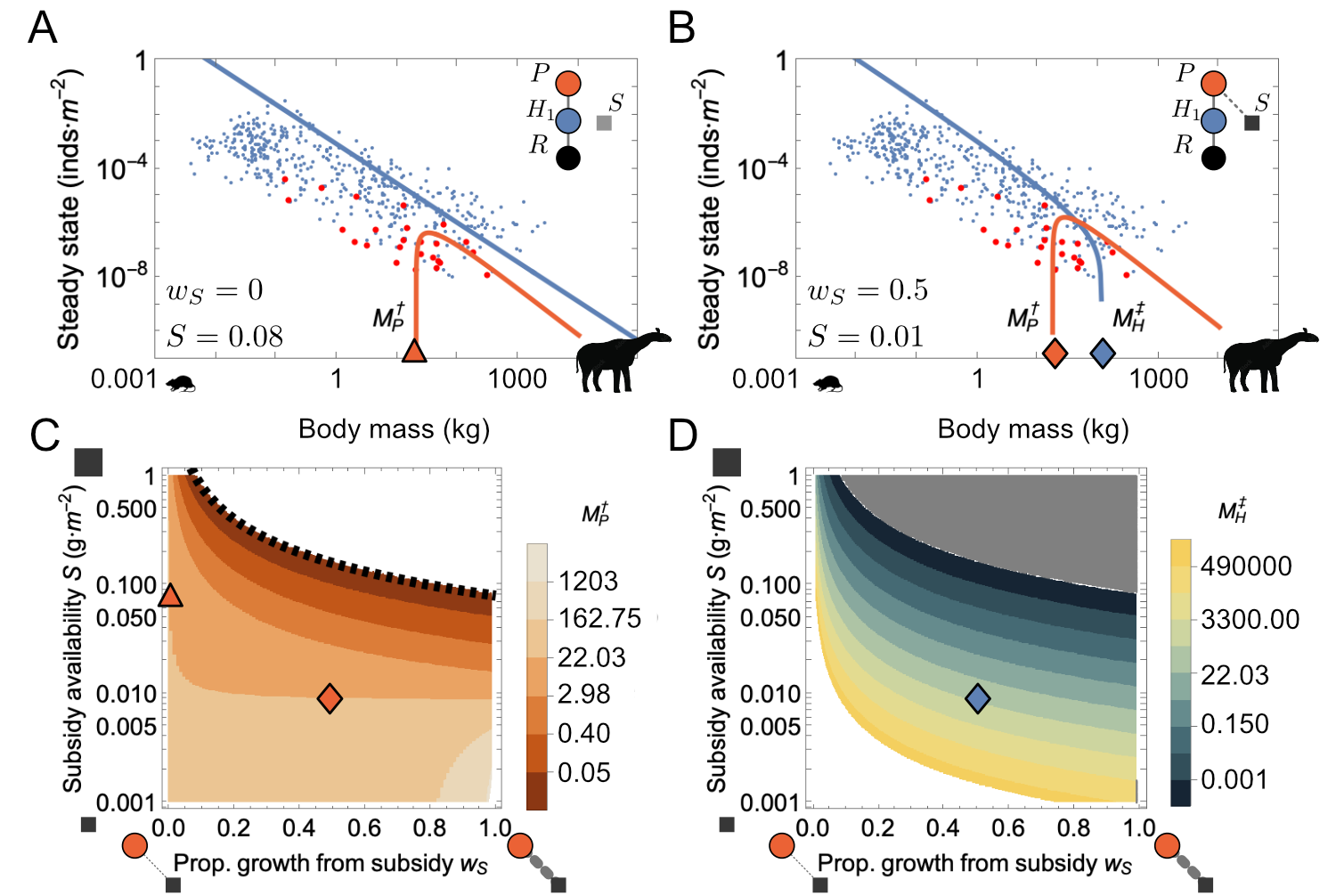}
    \caption{
        Population densities of the herbivore and predator in the subsidy model.
        (A) Blue and red lines denote the stable internal fixed points of the herbivore and predator populations, respectively, assuming 
        (A) no reliance by the predator on a subsidy ($w_S = 0$, $S=0.08)$, and 
        (B) with partial reliance by the predator on a subsidy ($w_S = 0.5$, $S=0.01)$.
        Blue and red points denote observed herbivore and carnivore densities, respectively \citep{Damuth1987,carbone2002common}.
        A low mass threshold for the predator $M_P^\dagger$ is denoted by a red triangle in (A); a low mass threshold for the predator, and a high mass threshold for the herbivore  $M_H^\ddagger$ are denoted by the red and blue diamonds, respectively, in (B).
        Predator low mass thresholds $M_P^\dagger$ (C), and herbivore high mass thresholds $M_H^\ddagger$ (D) as a function of the percent of predator growth driven by consumption of the subsidy $w_S$, and subsidy availability $S$.
        Black dotted line in (C) represents the critical value of subsidization $S_c$ enabling predator masses $0.01$ kg.
        White regions denote the absence of thresholds such that all size classes are feasible.
        Gray regions denote infeasibility for all size classes.
        Symbols in (A,B) are mapped to their corresponding positions in (C,D).
    }
    \label{fig:thresholdsub}
\end{figure*}
\endgroup

We observe specific limitations to the stability of the internal steady states for both the herbivore and predator as a function of their respective body mass and the dependence of the predator on external subsidization.
We define the transition from stable to unstable dynamics as the body mass where a population declines to zero and describe that point as a `body mass threshold'. 
These thresholds are defined by transcritical bifurcations, a feature observed in other allometric models of population dynamics -- typically at high-mass limits  \citep{weitz2006size,rallings2024}.
Throughout, we specify two distinct mass thresholds: a lower threshold (denoted by $\dagger$), below which populations are infeasible, and an upper threshold (denoted by $\ddagger$) above which populations are infeasible.

When the predator specializes on its primary prey, we find that there is a lower limit to predator body mass, $M_P^{\dagger}$, below which the population is infeasible such that the predator steady state vanishes (Fig. \ref{fig:thresholdsub}A).  
Here the smaller predator's reproductive output is stymied by its reliance on yet smaller prey, making the predator population unable to compensate for mortality unless externally subsidized.
Yet subsidization leads to the emergence of another feasibility threshold occurring at the upper-size limit of the herbivore prey population $M_H^{\ddagger}$, representing its maximum feasible body size (Fig. \ref{fig:thresholdsub}B).
As the predator's reliance on the subsidization is increased (moving along increasing $w_S$ in Fig. \ref{fig:thresholdsub}C, as long as $S$ is sufficient high), or as subsidy availability increases (moving  along increasing $S$ in Fig. \ref{fig:thresholdsub}C), we observe the lower mass threshold of the predator to decline. 
This means that small-predator body sizes above this declining threshold become feasible as the predator increasingly relies on the growing availability of subsidies.
The same changes have an opposing impact on the herbivore's large-size threshold due to the buoying effect subsidies have on the predator population, and the subsequent negative effects that predators have on herbivores (Fig. \ref{fig:thresholdsub}D).
As subsidy availability continues to increase, the large-size threshold of the herbivore population declines until all size classes are infeasible (gray region in Fig. \ref{fig:thresholdsub}D).

Our analysis also shows that there is a level of subsidization where the lower-mass threshold $M_P^\dagger$ is minimized, enabling the existence of very small predators.
This critical value of subsidization $S_c$ enabling predator masses of $0.01$ kg can be approximated as $S_c \approx S^{\rm int}w_S^{-1}$ where the intercept $S^{\rm int}=0.08$ g/m${}^2$ represents the subsidy required for a $0.01$ kg predator obtaining all of its growth from the subsidy to achieve positive population densities (dotted line in Fig. \ref{fig:thresholdsub}C).
The subsidies required to enable feasible populations at this lower-size limit become greater as the predator's reliance on subsidies decreases.

While extreme subsidization can decrease the small-mass threshold $M_P^\dagger$ such that effectively all predator body sizes are feasible, increased specialization on mammalian prey (lower $w_S$) tends to raise $M_P^\dagger$ (Fig. \ref{fig:thresholdsub}C).
For predators primarily reliant on herbivore prey, the small-mass threshold thus establishes a lower-bound on predator body size.
We can estimate the average minimum expected viable body size for a predator specializing on mammalian prey, averaged across values of $w_S<0.1$, as $\langle M_P^\dagger \rangle = 15$ kg, though the primary mode falls closer to $M_P^\dagger = 22$ kg (Appendix 3).

\subsection*{Competition and prey-switching}

The inclusion of two competing herbivores without external subsidization in the competition model, where $n=2$ and $S=0$ (Eq. \ref{eq:3d}), enriches the potential dynamics of the system, which vary with the relative body sizes of species as well as the flow of biomass between them.
This motif comprises a predator with body size $M_P$, the predator's primary herbivore prey with population density $H_1$ and mass $M_{H_1}={\rm E}\{M_H|M_P\}$, a secondary herbivore prey with population density $H_2$ and mass $M_{H_2} = \phi M_{H_1}$, and the basal resource with density $R$.
We note that the primary herbivore prey always has a body size following the expected prey size for a given predator mass, while the secondary prey can vary with $\phi$.
When $\phi<1$, the secondary prey is smaller than the primary prey; when $\phi>1$, the secondary prey is larger.
We note that $w_1$ is the proportion of predator growth from its primary prey, such that $w_2 = (1-w_1)$ reflects its reliance on the secondary prey; $w_1 \gg 0.5$ means that the predator is a specialist on its primary prey, $w_1\ll0.5$ means that the predator is a specialist on its secondary prey, whereas $w_1 \approx 0.5$ corresponds to predator generality.
Because of its status as primary prey, we may generally assume that $w_1>0.5$, though we show results for the full range.

The competition model has 13 fixed points, but only one stable internal fixed point for a particular combination of predator and herbivore masses (Fig. \ref{fig:propfeas}A).
As with the subsidy model, the stability of the system is similarly bounded by a predator low-mass threshold $M_P^\dagger$ and a primary prey high-mass threshold at $M_{H_1}^\ddagger$.
Importantly, this system additionally introduces a lower threshold for the secondary prey at $M_{H_2}^\dagger$.
In the example shown in Fig. \ref{fig:propfeas}A, where $w=0.8$ and $\phi=0.8$ (i.e., 80\% of the predators' growth is from its primary prey, which is 25\% larger than its secondary prey), predators with masses above the low-mass threshold for the secondary prey ($M_{H_2}^\dagger$) and below the high-mass threshold for the primary prey ($M_{H_1}^\ddagger$) have densities that are ratcheted to lower values.
This is due to the combined effects of competition and specialization on an increasingly scarce primary prey.
Immediately beyond the high-mass threshold for the primary prey, the predator population is buoyed by the secondary prey's release from competition.

\begingroup
\renewcommand{\baselinestretch}{1} 
\begin{figure*}[ht]
    \centering
    \includegraphics[width=0.95\linewidth]{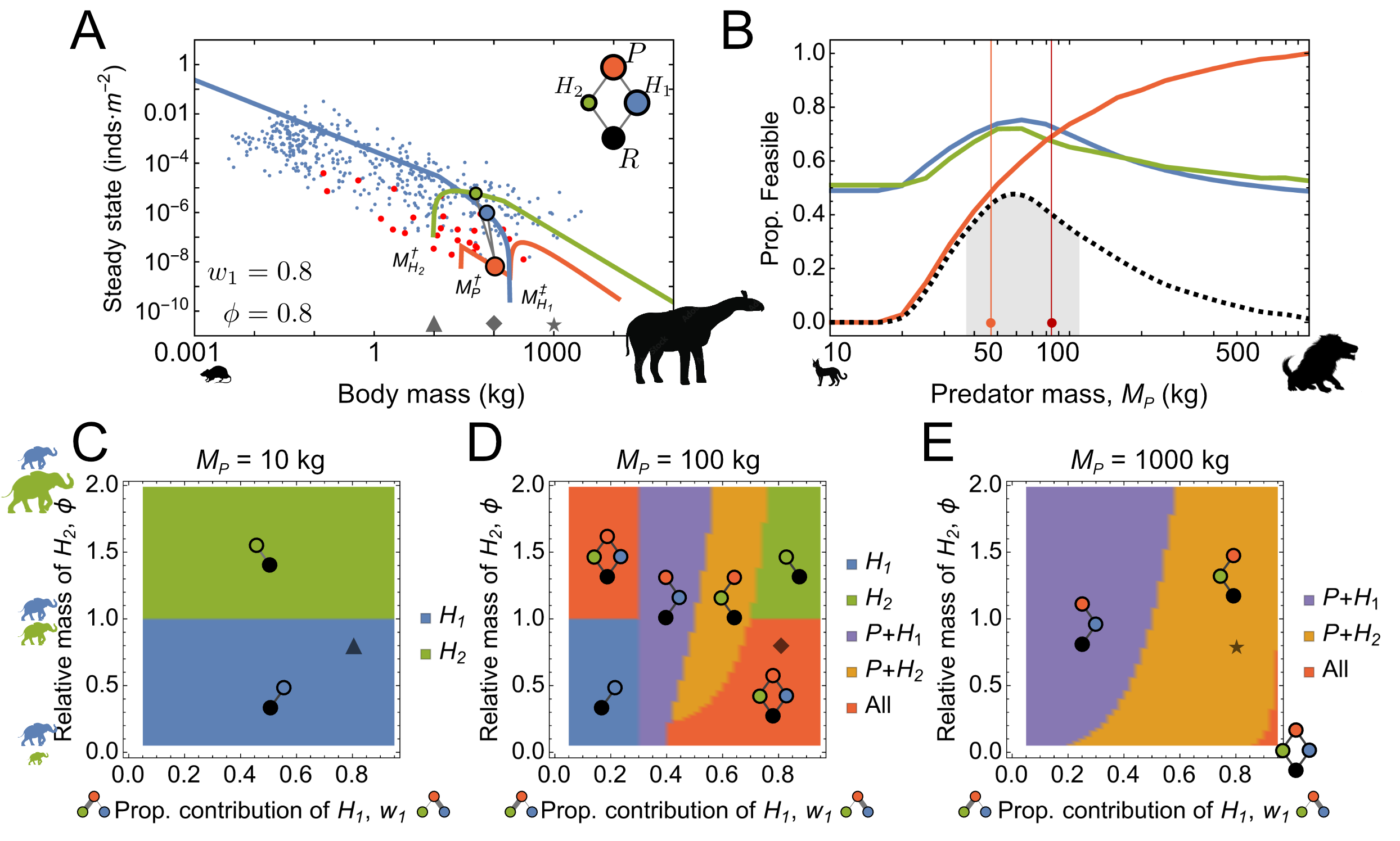}
    \caption{
        Population densities of species and the feasiblility of predators in the competition model.
        (A) Internal steady state densities for the predator and both herbivore prey, where the predator (red line) is specializing on the primary prey ($w_1=0.8$, blue line), where the secondary prey has a relatively smaller body size ($\phi=0.8$, green line).
        A predator-prey motif is highlighted by outlined points connected by trophic links on the density curves, where a 100 kg predator (red) consumes a ${\rm E}\{M_H|M_P\} = 74$ kg primary prey (blue) and a 59 kg secondary prey, given $\phi = 0.8$.
        Symbols along the body mass axis relate to coordinates in the coexistence regions shown in panels (C-E).
        (B) Proportion of parameter space $(w_1,\phi)$ resulting in feasible predator (red), and primary (blue) and secondary (green) prey  densities as a function of predator mass $M_P$.
        The predator mass where feasibility becomes zero is the predator lower mass threshold $M_P^\dagger=20$ kg.
        The proportion of parameter space that is feasible for all species is denoted by the stippled black line, and is maximized at $M_P=63$ kg, and where  $37<M_P<110$ kg (light gray shaded region) corresponds to a feasibility $>0.33$.
        Light and dark red points and vertical lines denote the geometric mean body size of contemporary and Cenozoic carnivores at $47$ and $84$ kg, respectively, taken across species above the threshold size of $20$ kg.
        (C-E) Coexistence of prey populations $H_1$, $H_2$, and the predator population $P$ as a function of predator body mass (C) $M_P = 10$ kg, (D) $M_P = 100$ kg, and (E) $M_P = 1000$ kg.
    }
    \label{fig:propfeas}
\end{figure*}
\endgroup

The number of coexisting species depends on the body sizes of predators and herbivore prey.
When the predator's mass is below the threshold $M_P^\dagger = 20$ kg, the predator population is not feasible, leaving the two herbivore populations to compete directly (Fig. \ref{fig:propfeas}B).
In this case, competition leads to exclusion of the smaller herbivore species, such that the herbivore population $H_1$ excludes $H_2$ when $\phi<1$, whereas $H_2$ excludes $H_1$ when $\phi>1$ (Fig. \ref{fig:propfeas}B,C).
As the predator mass $M_P$ approaches $100$ kg, the system straddles the lower-mass and higher-mass thresholds, where the coexistence of all species is possible (Fig. \ref{fig:propfeas}B,D).
In this region, predator specialization on the larger prey (high $w_1$ and low $\phi$, or low $w_1$ and high $\phi$) limits the larger herbivore's inherent fitness advantage, enabling coexistence of both herbivores with the predator (upper left and lower right corners of Fig. \ref{fig:propfeas}D).
The presence of all species in the motif defines a stable community, and the range of predator body masses where the proportion of feasible parameter space is greater than 33\% ranges from $M_P=37$ to $M_P=110$ kg (shaded region in Fig. \ref{fig:propfeas}B; peak feasibility at $M_P=63$ kg).
If the predator specializes on the smaller prey, both of these populations collapse, leaving the larger prey to persist alone (lower left and upper right corners of Fig. \ref{fig:propfeas}D).

In contrast to these dietary extremes, predator generalization ($w\approx 0.5$) results in exclusion of the prey that contributes to a slightly larger proportion of predator growth, except when $\phi$ is low, in which case all species can coexist.
For higher values of $\phi$, this pattern is largely insensitive to the relative size of the secondary prey (along the $\phi$ axis; purple and yellow areas in Fig. \ref{fig:propfeas}D-E).
Here the larger-bodied predator has an expected prey with similarly larger body size, but where the lower reproductive rates of these larger prey cannot shield against the effects of a predator obtaining a significant portion of its growth from the secondary prey.
This effect is similar to that of increased predator subsidization, lowering the herbivore upper threshold $M_H^\ddagger$ and increasing the range of herbivore body sizes that are infeasible.
It is thus the larger prey that is more likely to become the victim of increased predator subsidies, whether they arise from external resources or a secondary prey.
This pattern expands across all $(w,\phi)$ as predators reach megapredator size classes ($M_P \rightarrow 1000$ kg; Fig. \ref{fig:propfeas}E). 
The asymmetry in the boundaries between these steady states is likely due to \emph{i}) the super-linear scaling of ${\rm E}\{M_H|M_P\}$, and \emph{ii}) the disproportionate spacing between $\phi > 1$ and $\phi < 1$, where the range of possible secondary prey masses is compressed below $\phi = 1$ and extended above it.

\subsection*{Predator dietary generality}

While there are many ways to define dietary generality, here we adopt a relatively simple dichotomy, where we define a generalist predator -- in the context of the competition model -- as one with a dietary breadth consisting of an even contribution of two prey ($w_1 \approx 0.5$), whereas specialization increases with the contribution of a single prey. 
We emphasize that this definition highlights the relative importance of fueling growth from multiple mammalian prey of potentially different body sizes.
The relative advantage of generalist versus specialist diets, as a function of predator body size, can be evaluated by comparing the steady state densities of both.
We quantify the predator's generalist advantage as the ratio of generalist to specialist steady state densities $P^*_{\rm Gen}/P^*_{\rm Spec}$ for a predator of a given body mass, where we define generality by the condition $w_1=0.5$, moderate specialization on the primary prey by the condition $w_1=0.75$, and extreme specialization on the primary prey by the condition $w_1=0.85$, where we assume that the secondary prey is smaller than the primary prey ($\phi=0.2$).
As such, a value of $P^*_{\rm Gen}/P^*_{\rm Spec} > 1$ means that the generalist dietary strategy benefits the predator population, while $P^*_{\rm Gen}/P^*_{\rm Spec} < 1$ means the specialist strategy is more beneficial.
We observe that, in parameter regions where predators are feasible, there is a generalist advantage, relative to moderate specialization, for predator size classes $57 < M_P < 420$ kg (dark red line, Fig. \ref{fig:preddens}A).
This advantage extends to larger size classes when generalization is compared against extreme specialization, where the advantage declines at ca. $668$ kg (light red line, Fig. \ref{fig:preddens}A).

Our results show that the generalist advantage, with respect to either form of specialization, declines as predator body size increases, giving way to a megapredator specialist advantage (Fig. \ref{fig:preddens}A,C).
In this specialist-advantage range we observe that the megapredator drives its primary prey to extinction, subsisting instead on the smaller and more abundant secondary prey (Fig. \ref{fig:propfeas}E).
This effect is roughly symmetrical about $w_1$ and qualitatively invariant along $\phi$, meaning that moderate specialization on either the primary or secondary prey, regardless of the size difference between them, promotes the predator's steady state population density (Fig. \ref{fig:preddens}C).

\begingroup
\renewcommand{\baselinestretch}{1} 
\begin{figure*}[ht]
    \centering
    \includegraphics[width=0.95\linewidth]{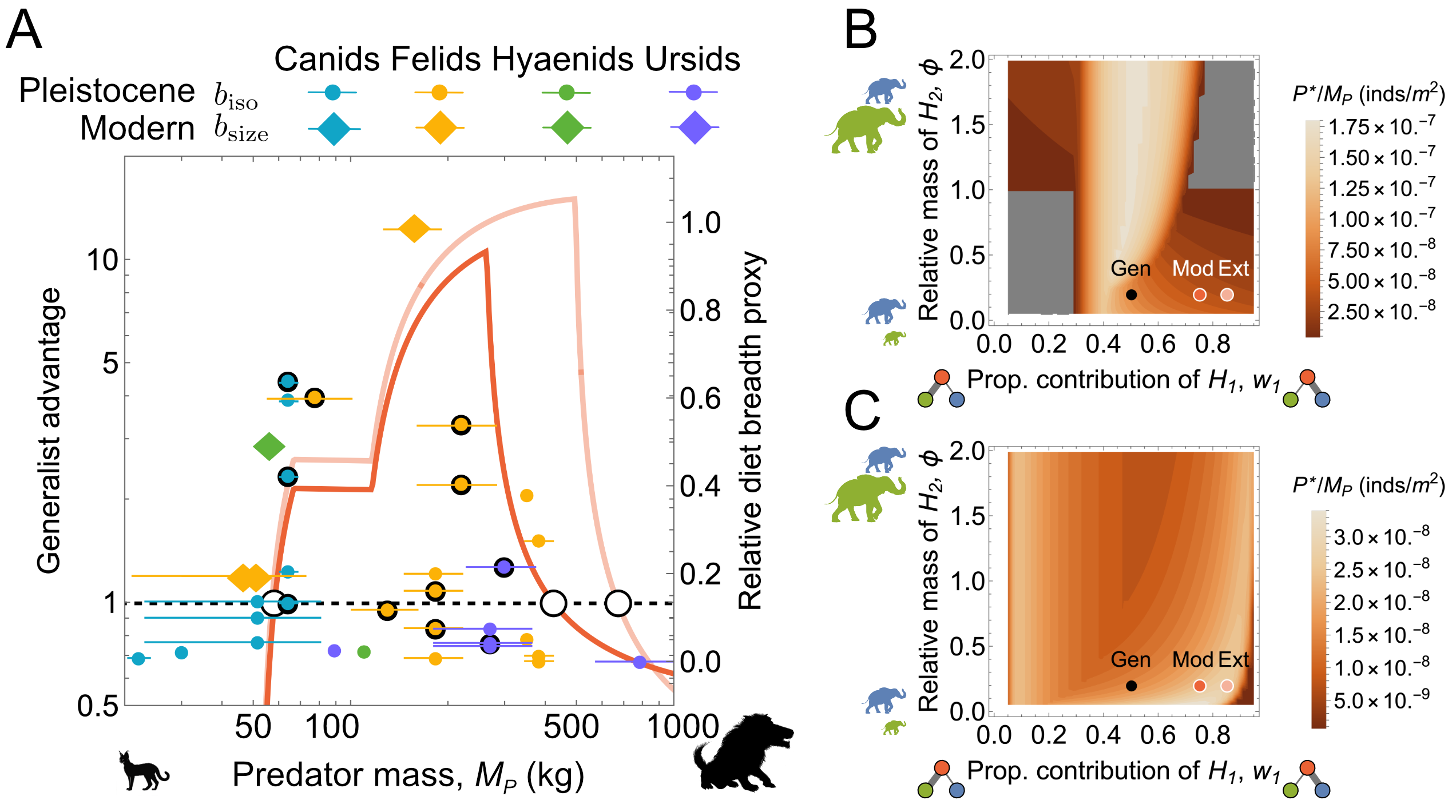}
    \caption{
        (A) The ratio of generalist ($w_1=0.5$) predator densities to moderate specialist ($w_1=0.75$; dark red line) and extreme specialist ($w_1=0.85$; light red line) predator densities ($P^*_{\rm Gen}/P^*_{\rm Spec}$; left y-axis) as a function of predator mass $M_P$, where the relative size of the secondary prey is smaller than the primary prey ($\phi=0.2$).
        A generalist advantage exists for values of $P^*_{\rm Gen}/P^*_{\rm Spec} > 1$.
        White circles along the dashed line denote where $P^*_{\rm Gen}/P^*_{\rm Spec} = 1$.
        Colored points denote Pleistocene (small circles) and contemporary (large diamonds) relative diet breadth proxies ($b_{\rm iso}$ and $b_{\rm size}$) as a function of predator size for canids (teal), felids (orange), hyaenids (green), and ursids (purple).
        Black outlined colored circles denote $b_{\rm iso}$ calculated from carbon isotopic range, while non-outlined colored circles denote values calculated from carbon and nitrogen isotopic convex hull area.
        (B) and (C) Predator steady state densities $P^*/M_P$ (inds/m${}^2$) across $(w_1,\phi)$ for 
        (B) large ($M_P=100$ gg) and 
        (C) megapredators ($M_P = 1000$ kg).
        Black points represent the generalist and red/orange points represent the moderate (Mod) and extreme (Ext) specialist values used in (A). 
        Gray regions denote predator infeasibility for all size classes.
    }
    \label{fig:preddens}
\end{figure*}
\endgroup

To determine whether our predictions of the generalist advantage accord with estimates of dietary generality from both contemporary and extinct large-bodied ($> 20$ kg) mammalian predators, we examine two separate proxies of relative dietary breadth.
We specifically focused on predators specializing on vertebrate prey, and do not include contemporary omnivores with observed diets diverging from vertebrate specialization.
For extant predators, which include Serengeti cheetah, leopards, hyaena, and lions, dietary breadth is measured directly as the ratio of the prey weight range consumed by each predator relative to the prey weight range available to all predators \citep{Sinclair2003}.
In this case, relative dietary breadth is given as $b_{\rm size} = q_{\rm pred}/q_{\rm herb}$, where $q_{\rm pred}$ is the prey mass range utilized by individual predator species, and $q_{\rm herb}$ is the prey mass range available to all predator species.
For predators consuming a larger range of prey body sizes, $b_{\rm size} \rightarrow 1$, implying dietary generalization.
We observe this proxy of diet breadth to increase in good alignment with the predicted generalist advantage (Fig. \ref{fig:preddens}A).
The smaller-sized cheetahs and leopards have a limited prey size range, whereas hyenas and lions have increasingly larger prey size ranges, corresponding to the peak generalist advantage relative to moderate specialization.

Because contemporary mammalian predators are not representative of the full size range realized by predators in the fossil record for millions of years until very recently, we include 33 additional predator species from 14 separate assemblages spanning the early Pleistocene to the early Holocene, including dire wolves (\emph{Aenocyon dirus}), American lions (\emph{Panthera atrox}), saber-toothed cats (\emph{Smilodon} spp.), and the short-faced bear (\emph{Arctodus simus}), which is the largest included mammalian predator at ca. $780$ kg \citep{palmqvist2008tracing,fox2008pleistocene,koch2004effects,desantis2021dietary,trayler2015inland,fuller2020pleistocene,fuller2014ultrafiltration,coltrain2004rancho,feranec2014understanding}. 
While short-faced bears are thought to have been vertebrate specialists \citep{FoxDobbs:2008tq}, included in our Pleistocene dataset are some smaller-bodied ursids that may have been more omnivorous.
For these extinct species, dietary breadth is estimated from carbon isotope values (measured as $\delta^{13}{\rm C}$) extracted from bone collagen or tooth enamel apatite for both predators and their potential herbivore prey, specific to each assemblage (see Appendix 5).
Relative diet breadth is then estimated as $b_{\rm iso} = r_{\rm pred}/r_{\rm herb}$, where $r_{\rm pred}$ is the maximal isotopic range of each predator species, and $r_{\rm herb}$ is the maximal isotopic range of all available herbivore species in a given assemblage. 
As such, if predators rely on the full suite of available prey (to different extents, among individuals and over time), their isotopic range is more likely to approximate that of the available herbivore community, such that $b_{\rm iso}\rightarrow1$.
On the other hand, if predator individuals specialize on a particular herbivore species or set of species, it is more likely that $b_{\rm iso} < 1$.

We observe that the isotopic proxy of diet breadth reveals much greater variability among extinct carnivores, though this variability is roughly contained within the predicted zone of generalist advantage.
At smaller size classes, we observe carnivores to have lower isotopic breadth, increasing for dire wolves and the similarly-size \emph{Smilodon gracilis} along the predicted trajectory of the generalist advantage.
At larger size classes, felids including other \emph{Smilodon} spp., the American lion, and \emph{Homotherium} spp. fill out the range of specialization to generalization.
In contrast, ursids occupy a relatively smaller proportion of isotopic space implying comparative specialization, including the only megapredator representative, the short-faced bear.

\section*{Discussion}

The effects of bioenergetic trade-offs on the density and, by extension, feasibility, of a species' population depends on its body size alongside the body sizes of those species in its local network of interactions \citep{rallings2024,brose2010body}.
Although such trade-offs are known to structure the larger community through universal predator-prey biomass scaling laws \citep{Hatton:2015fk}, the mechanisms linking individual bioenergetics to community-level patterns remain elusive.
We argue that bioenergetic bounds on population feasibility play a central role in shaping the fitness landscape contributing to larger-scale trends in body size evolution.

The large shifts in mammalian body size distributions following the K-Pg extinction were initiated by competitive release and accelerated by post-extinction climate change \citep{lyson2019exceptional}. 
As body size is a primary determinant of ecological interactions \citep{weitz2006size}, it is reasonable to assume that interactions played an important role in shaping subsequent size-evolution throughout the Cenozoic, establishing the foundation for the structure and function of contemporary mammalian communities. 
Our theoretical framework allows us to identify and locate bioenergetic population instabilities emerging from interactions such as predation and competition as a function of body size.

\subsection*{Subsidies, limits to predator specialization, and megaherbivore collapse}

External dietary subsidies apart from traditional prey can arise from a variety of sources, including alternative trophic levels (e.g. omnivores), carcasses \citep{Wilson:2011bl,ritwika2024beyond}, adjacent biomes \citep[e.g. marine subsidies;][]{darimont2009landscape,pires2023beyond}, and more recently anthropogenic activities \citep{hopkins2014changing}. 
On average, the majority of observed mammalian diets are accounted for by two different prey taxa, with potentially a range of others contributing the remainder \citep{Hutchinson2022}.
While some subsidies (e.g., alternative prey) may be impacted by consumption, the availability of other sources may be large enough (e.g., highly abundant energy sources), or separated enough (e.g., transfer between ecosystems), that there is not a dynamic feedback between the subsidy and the beneficiary.
Our subsidy model ($n=1$, $S>0$; Eq. \ref{eq:3d}) assumes the latter, where both a constant subsidy density $S$, as well as the proportional reliance of the predator's growth on the subsidy $w_S$, must be specified.

Our model predicts that a predator population specializing on a mammalian herbivore as prey is subject to a mass-specific instability at body sizes $M_P\leq16-22$ kg.
Importantly, this finding directly supports the observation that vertebrate-feeding predators smaller than ca. 20 kg must subsidize their diet with additional resources \citep{Carbone:1999ju}, and that increasing subsidy richness $S$ is required to enable ever smaller predator body sizes (lower $M_P^\dagger$; Fig. \ref{fig:thresholdsub}A).
\citet{Carbone:1999ju} predicted a threshold for vertebrate-specialist carnivores of ca. 20 kg based on the individual energetic limitations associated with hunting vertebrate prey as a function of predator body mass.
This predicted size limit corresponds to an observed change in predator dietary requirements: those below the limit have diets with significant contributions of non-vertebrate resources, while those above tend towards vertebrate specialization.
Our framework makes a similar prediction, but one emerging from a population instability driven by bioenergetic trade-offs rather than individual foraging energetics, suggesting our model captures fundamental features of natural systems.

The subsidization of predators has large implications for the survival of their herbivore prey.
As has been previously demonstrated, increasing subsidy densities serves to promote predator growth, ultimately driving herbivore populations to collapse \citep{nevai2012effect,Holt:2008wa}.
In this regime, there is not a cascading effect on the predator population because they are buoyed by the external subsidy.
With subsidization, an upper threshold emerges for the herbivore prey (Fig. \ref{fig:thresholdsub}B), serving as an upper body size limit.
As subsidization increases, the upper threshold declines, limiting the range of feasible body sizes (Fig. \ref{fig:thresholdsub}B).
This serves to destabilize the largest megaherbivores at the lowest subsidy densities, ultimately eliminating herbivore feasibility if subsidies to the predator increase enough (gray region, Fig. \ref{fig:thresholdsub}B). 
We suggest that the implicit connection between the size-specific stability of herbivore populations to predator subsidization may have direct implications for conservation planning, as predator access to subsidies directly impacts prey viability \citep{pires2023beyond}. 
Combining the approach we use here with tools from population viability analysis \citep{doak2009population} may enable more robust estimates informed by energetic trade-offs.  

Despite the physiological and ecological advantages associated with larger body size \citep{Dunbrack:1993cs,Smith:2011cza}, there is also considerable risk.
This risk is largely due to the smaller population densities of larger-bodied species \citep{Brook2005,ripple2010linking}, as well as the significantly longer timescales of gestation \citep{jackson2014basal}.
These factors suggest that populations of larger-bodied species are both more prone to collapse, take longer to recover if they have the opportunity \citep{owen1988megaherbivores,ripple2015collapse}, with disproportionately large cascading impacts on ecosystem functioning \citep{enquist2020megabiota,Hempson:2015hk}.
Over evolutionary timescales, these risks may contribute to shorter temporal ranges for larger-bodied clades \citep{Liow:2008jx}. 
While the determinism of these relationships is not always clear, the well-documented end-Pleistocene mass extinction reveals demonstrable size-selectivity \citep{Smith:2018gm,Koch:2006vt}.
Our model clearly supports size-selective destabilization when predators are subsidized, as would be the case for Pleistocene hunting and gathering \emph{Homo sapiens} \citep{thompson2019origins,bradshaw2024small}.

\subsection*{Fasting endurance and predator size evolution}

Increasing the diversity and complexity of ecological communities can alter the outcome of interactions \citep{spaakbuildingmodern2023,Chesson:2008id}. 
We observe that our competition model, which includes a predator and two competing herbivore prey ($n=2$, $S=0$; Eq. \ref{eq:3d}), substantially increases the dynamical richness of the system.
The emergence of three distinct body size thresholds (Fig. \ref{fig:propfeas}A) -- an upper threshold for the primary prey ($M_{H_1}^\ddagger$), and lower thresholds for the predator ($M_{P}^\dagger$) and secondary prey ($M_{H_2}^\dagger$) -- indicate that alternative community structures depend on specific combinations of species' masses and biomass flow (Fig. \ref{fig:propfeas}C-E).
These thresholds, which vary with dependence of the predator on each prey ($w_1$) and the relative size difference between prey ($\phi$), combine to determine which species in the interaction motif can maintain positive population densities.
While the positions of these mass thresholds change with the size-difference of herbivores ($\phi$) and the relative reliance of the predator on the primary prey ($w_1$), the presence of these thresholds is generally robust (Fig. \ref{fig:propfeas}C-D), and appear even when $w_1$ is treated as a dynamic variable (Appendix 4).

As with the subsidy model, we find that smaller-bodied mammalian-specialist predators (ca. $M_P<20$ kg) cannot maintain viable population sizes, even with multiple prey.
As the predator is not externally subsidized in the competition model, this appears to be a hard limit (Fig. \ref{fig:propfeas}B).
This reinforces the notion that there is a predator size threshold at ca. 20 kg that serves as a stability boundary for mammalian-specialization below which integration of non-mammalian resources is required, and supporting the conclusions of the energetic model proposed by \citet{Carbone:1999ju}.
Smaller predators that are not viable become extinct, leaving the herbivores -- which vary only in body size and the associated scaling of their vital rates -- to compete with one another.
As a result, the larger herbivore excludes the other given its increased starvation tolerance and lower sensitivity to declines in resource abundance (Fig. \ref{fig:propfeas}C) \citep{yeakel2018dynamics,rallings2024}, mirroring observations of positive size-selectivity resulting from resource competition \citep{brown1986body,Bonner1988,Kingsolver2004}.

The role of herbivore body mass in determining the outcomes of interspecific competition is complex \citep{sinclair1985does,murray2000vegetation,arsenault2002,illius1992modelling}, challenged by alternative arguments based on dietary differentiation across plant taxonomic groups \citep{Kartzinel2015,Pansu2022} and tissues \citep{Jarman1979}, coupled with detailed descriptions of how vegetation structure influences foraging behavior \citep{ritchie1999spatial,shipley2007influence,bhat2020scaling} and interspecific interactions \citep{bell1971grazing}.
On the one hand, smaller herbivores have lower absolute food requirements and can be more selective in their foraging behaviors \citep{shipley2007influence,clauss2013}. 
On the other hand, larger herbivores can tolerate lower quality foods due to lower mass-specific metabolic requirements \citep{clauss2013,Dunbrack:1993cs}, and their sheer size reduces their vulnerability to predation \citep{Sinclair2003}.
One aspect of size-based competition among herbivores that is rarely considered is the superlinear scaling of fat-storage capacity with body mass \citep{CalderIII:1983jd}, which confers greater starvation resistance in larger species assuming they can utilize this capacity \citep{Millar:1990p923}. 
The impact of greater starvation resistance in large mammals and the competitive advantage it confers over smaller species during temporary periods of extreme food scarcity may contribute to the selection for larger body sizes known as Cope's Rule \citep{yeakel2018dynamics}.

\citet{brown1993evolution} demonstrated that an optimal body size emerges from balancing the rates of energy acquisition and conversion into reproductive power at the individual level.
Although such energetic trade-offs explain large-scale features of body size evolution, considering species interactions allows us to pinpoint specific body size thresholds within ecological communities.
Our framework emphasizes the combined roles of trophic interactions and starvation resistance at the population level in shaping a range of predator body sizes that serve to promote feasible population densities of the local community. 
We observe that at $M_P \approx 100$ kg, coexistence of all species is possible as long as the predator specializes on the largest herbivore (Fig. \ref{fig:propfeas}D), which serves to suppress the latter's inherent competitive advantage.
Across parameter values for $(w_1,\phi)$, the probability that all species in the tri-trophic competition motif coexist exceeds 30\% ranges from $M_P=37$ to $M_P=110$ kg (shaded region in Fig. \ref{fig:propfeas}B), with the peak corresponding to a predator size roughly that of a spotted hyena ($M_P=63$ kg).
Predator and herbivore body sizes occupying this region tend to fall between the lower- and upper body mass thresholds (Fig. \ref{fig:propfeas}A). 

Our prediction of a predator mass giving rise to a maximally stable trophic structure -- while specific to the competition model -- aligns with the contemporary large-bodied terrestrial carnivore body size average, with a geometric mean of ca. $47$ kg (light red point in Fig. \ref{fig:propfeas}B) \citep{Smith:2004p1595}, accounting for only terrestrial carnivores above the established 20 kg vertebrate-specialization limit \citep{Carbone:1999ju}.
Notably, the contemporary geometric mean is less than that for terrestrial Carnivora across the Cenozoic at $84$ kg (dark red point in Fig. \ref{fig:propfeas}B) \citep{smith2003body}.
We suggest that this alignment between model prediction and observational data may point to a potential `attractor' for mammalian predator body sizes, where the feasibility of a local neighborhood of interacting species is optimal.
This body size range could be viewed as a stable core fueling the hypercarnivore ratchet described by Van Valkenburgh et al. \citep{VanValkenburgh:2004p2451}, where short-term success attained by evolution beyond this core size range is followed by long-term failure, and where the antecedents emerge again from this core to fuel the next tooth in the ratchet.

The dynamics that we observe may additionally shed light on specific drivers of predator size-evolution beyond this so-called stable core.
Our results show that stability of the motif is maintained only if the predator specializes on the larger of the two herbivore prey (red regions in Fig. \ref{fig:propfeas}D), highlighting a potential fitness gradient.
As the stability of the predator population is promoted by the consumption of larger-bodied prey, so would those traits (such as larger predator size) that increase the likelihood of successful acquisition and handling.
Such positive size-selection of the predator may, in turn, contribute to positive size-selection within targeted herbivore clades to escape predation \citep{Benton2002}, adding fuel to the established fitness advantages associated with starvation resistance and promoting a coevolutionary response. 
At mega size-classes, such an arms race would eventually hit the large-mass herbivore threshold $M_H^\ddagger$, perhaps an evolutionary expression of self-organized criticality \citep{sole1996extinction}, where the herbivore prey is squeezed between the dual pressures of lower reproductive output and mortality induced by a predator large enough to specialize on it.
Our framework indicates that this increased predation pressure would serve to lower the upper mass threshold of the herbivore prey, such that $M_H^\ddagger < M_{H}$, resulting in its collapse with potentially cascading effects. 

\subsection*{Adaptive benefits of predator dietary generality}

Mammalian carnivores tend to incorporate a larger diversity of prey as they increase in body size \citep{gittleman1985carnivore}, fueled in part by their access to larger spatial areas \citep{Garland:1983va} and enhanced abilities to subjugate a larger range of prey \citep{Carbone:2007dz,Sinclair2003}.
This has significant consequences for the structure of species interactions, contributing to nested patterns of interaction in contemporary terrestrial communities \citep{Baskerville:2010vn,Olff:2009tk}.
We evaluate the potential advantages of generalist versus specialist feeding directly, by comparing the relative densities of predators where we set $w_1=0.5$, $w_1=0.75$, and $w_1=0.85$ to represent generalist, moderate specialist, and extreme specialist feeders, respectively.
Here we hold the secondary prey to be smaller than the primary prey ($\phi=0.2$) to ensure that the full tri-trophic motif can persist at steady state (see Fig. \ref{fig:propfeas})D), regardless of generalization or specialization. 

Relative to moderate specialists, we observe increasing advantages of generalist-feeding between predator body sizes $M_P=57$ kg and $M_P=421$ kg (Fig. \ref{fig:preddens}).
Above this size range, we find that it is more advantageous to specialize.
If the generalist advantage is compared to extreme specialists, the relationship reaches this change-point at an increased $M_P=668$ kg.
This suggests that the advantage of being a generalist predator diminishes at very large body sizes, where specialization becomes more beneficial. 
Perhaps compellingly, if we remove the assumption of a constant $w_1$, instead allowing it to vary dynamically with the relative abundances of prey, we obtain independent support for this finding.
Assuming dynamically changing $w_1$, we observe increasingly large transient oscillations in herbivore population densities as generalist predator body sizes exceed 600 kg, increasing the risk of collapse.
In contrast, specialist megapredators cause comparatively smaller transient oscillations in their prey populations (see Appendix 4).
The alignment of the predicted megapredator specialist advantage presented in Fig. \ref{fig:preddens}, and that revealed by allowing $w_1$ to vary dynamically, provides additional support for the notion that megapredator diets may have been constrained to specialization compared to their smaller-bodied conspecifics, which includes the largest terrestrial predators in contemporary systems.

The expected generalist advantage closely aligns with observations of prey size ranges for contemporary carnivores \citep{Sinclair2003}, where our metric of relative dietary breadth ($b_{\rm size}$) increases sharply with hyenas and lions along the curve denoting the generalist advantage (diamonds, Fig. \ref{fig:preddens}A).
The incorporation of an isotopic proxy for dietary breadth ($b_{\rm iso}$) allows us to evaluate the potential alignment of Pleistocene carnivores, capturing a fuller picture of terrestrial megafaunal trophic interactions.
Because $b_{\rm iso}$ is a rough proxy of diet breadth, it is prone to specialization bias.
For instance, a smaller isotopic range or area may include many species of herbivore (e.g. multiple grazers consuming ${\rm C_4}$-photosynthetic grasses), such that a predator generalizing on many grazing herbivores may have a narrow isotopic distribution that incorrectly prescribes a specialist dietary niche.
The reverse bias is less likely to occur, given that a predator isotopic spread on the order of that of the herbivore community is more likely to be a stronger signal of dietary generalization.
The distribution of $b_{\rm iso}$ values clearly demonstrates an increased potential for generalist dietary strategies among larger-bodied carnivores, including saber-toothed cats (\emph{Smilodon} spp.) and the American lion (\emph{Panthera atrox}), although these and other large-bodied species across multiple assemblages run the gamut of dietary proclivities, including extreme specialization.
Of particular note is the largest carnivore, the short-faced bear (\emph{Arctodus simus}), and while it is only a single datum, it aligns with the predicted decline of the generalist advantage at megapredator size classes.

Contemporary mammalian communities may engender a biased perspective on the form and function of past ecosystems \citep{sondergaard2025}.
Alongside the heightened diversity and larger sizes of megaherbivores in past ecosystems \citep{faith2018plio,Faith2019}, larger predators also reached mega-size classes.
Megapredators, including the mid-Eocene \emph{Andrewsarchus} (an artiodactyl), the late Eocene \emph{Sarkastodon} (an oxyaenidont), and the mid-late Miocene \emph{Amphicyon ingens} (a bear-dog), have estimated sizes ranging from 550 to 1000 kg \citep{sorkin2008biomechanical,burness2001dinosaurs}.
These species are perhaps outliers to the mammalian predatory condition, though species of more recent genera such as \emph{Arctodus} and \emph{Smilodon} have estimated sizes nearing 500-850 kg \citep{sorkin2006ecomorphology,christiansen1999size,christiansen2005body,anyonge1993body}, so perhaps this condition is not so rare as our contemporary ecosystems would suggest.
While little direct evidence exists for the diets of these and other mammalian megapredators, the largest felids reveal variable dependence on both grazing and browsing herbivores \citep{feranec2004isotopic,desantis2019,Yeakel:2012uc}, while isotopic evidence indicates short-faced bears were largely specializing on caribou \citep{FoxDobbs:2008tq,Yeakel:2012uc}.
Our theoretical framework predicts that species including saber-toothed cats and the American lion would be capable of generalizing to a similar extent, if not more than, contemporary lions, with perhaps only the largest megapredators such as \emph{Sarkastodon} and \emph{Andrewsarchus} reverting to specialization.
This accords with biomechanical arguments, where it is thought that the extreme size of these megapredators would limit their potential prey to only the slowest and largest megaherbivores \citep{sorkin2008biomechanical}.

\subsection*{Conclusions}

Understanding macroevolutionary pressures through the lens of ecological interactions provides inherent value, where the selective forces that shape and reshape the long-term dynamics of communities can be elucidated.
We suggest that because our framework is premised on fundamental bioenergetic trade-offs, it may be useful for both reconstructing past ecological changes in response to known environmental upheaval, as well as forecasting future risks. 
As we face this `age of emergency' \citep{byrne2021actual} characterized by accelerating anthropogenic impact, environmental disturbance, and a slide into no-analog conditions, it is imperative to harvest insight from the patterns and processes that have shaped mammalian diversity throughout the Cenozoic.

\section*{Acknowledgments}
We would like to thank Irina Birskis-Barros, Jessica Blois, Nathaniel Fox, Paulo Guimar\~aes Jr., and Emily Lindsey for insightful comments and discussions that greatly improved the ideas and concepts that contributed to this manuscript. These ideas benefited greatly from travel funds provided to JDY from the Santa Fe Institute. This project was supported by National Science Foundation grant EAR-1623852 to JDY.


\end{bibunit}


\setcounter{secnumdepth}{2}

\renewcommand{\thesection}{Appendix \arabic{section}}
\renewcommand{\thesubsection}{Appendix \arabic{subsection}}
\renewcommand{\thefigure}{S\arabic{figure}}
\renewcommand{\theequation}{S\arabic{equation}}

\setcounter{figure}{0}
\setcounter{equation}{0}


\clearpage

\begin{bibunit}

\section*{Supplemental Appendices}

\subsection{Allometric timescales for vital rates}
\label{sec:timescales}

The rate laws describing the growth and mortality of both predator and consumer vary as a function of predator body mass $M_P$ and consumer body mass $M_C$, representing a mammalian carnivore and herbivore, respectively. 
We approach the derivation of vital rates with respect to predator and consumer mass by solving for multiple timescales associated with ontogenetic growth, maintenance, and expenditure.
The growth of an individual predator or consumer $i$ from birth mass $m_i=m_0$ to its reproductive size $m_i=0.95M_i$ (where $M_i$ is its observed adult mass) is given by the solution to the general balance condition $B_0 m_i^\eta = E_m \frac{\rm d}{\rm dt}{m_i} + B_m m_i$, where $E_m = 5774~{\rm J}\cdot{\rm g}^{-1}$ is the energy needed to synthesize a unit of biomass \citep{moses2008rmo,hou2008}, $B_0 = 0.047~{\rm W \cdot g}^{-\eta}$ is the metabolic normalization constant \citep{Pirt1965}, $B_m$ is the metabolic rate to support an existing unit of biomass, and the metabolic exponent $\eta=3/4$ \citep{West:2001bv}.
Birth size is assumed to follow the allometric relationship $m_0 = 0.097 M^{0.92}$ \citep{blueweiss1978relationships,stryer}.
From this balance condition, the time required for an organism starting from mass $m^\prime_i$ to reach mass $m^{\prime\prime}_i$ follows
\begin{equation}
    \tau(m_i^{\prime},m_i^{\prime\prime}) = \ln\left(\frac{1 - (m_i^{\prime}/M_i)^{1-\eta}}{1-(m_i^{\prime\prime}/M_i)^{1-\eta}}\right)\frac{M_i^{1-\eta}}{a(1-\eta)},
\end{equation}
where $a = B_0/E_m$.
From this general equation, we calculate the timescale of reproduction for a predator or herbivore consumer of mass $M_i$ as $t_{\lambda_i} = \tau(m_0,0.95M_i)$, such that the reproductive rate is $\lambda^{\rm max}_i = \ln(\nu)/t_{\lambda_i}$, where $\nu=2$ is the set number of offspring per reproductive cycle \citep{Savage2004,yeakel2018dynamics}.
Yield coefficients for the predator and consumer are given by $Y_i = M_i E_d/B_{\lambda_i}$ where $B_{\lambda_i}$ is the lifetime energy use required to reach maturity $B_{\lambda_i} = \int_0^{t_{\lambda_i}} B_0 m_i(t)^\eta {\rm dt}$, and $E_d$ is the energy density of that which is being consumed \citep{yeakel2018dynamics}.

To determine the maximum rate of mortality due to starvation $\sigma_C^{\rm max}$, we calculate the time required for a consumer to metabolize its endogenous energetic stores, estimated from its cumulative fat and muscle mass \citep{yeakel2018dynamics,rallings2024}.
During starvation, we assume that an organism burns its existing endogenous fat stores as its sole energy source, such that the balance condition is altered to $\frac{\rm d}{\rm dt}{m_C}E^\prime_m = -B_m m_C$, where $E_m^\prime = 7000~{\rm J}\cdot{\rm g}^{-1}$ is the amount of energy stored in a unit of biomass \citep{Pirt1965,Dunbrack:1993cs}, which differs from the amount of energy used to synthesize a unit of biomass $E_m$ \citep{yeakel2018dynamics}.
The natural mortality rate $\mu_i$ is derived by assuming that population cohorts decline over time following a Gompertz relationship \citep{CalderIII:1983jd}.
Using mammalian allometric relationships for the average initial cohort mortality rate, the actuarial mortality rate, and organismal lifespan, we obtain $\mu_i$ as a function of body mass $M_i$ \citep{rallings2024}.

The natural mortality rate is obtained by first assuming that the number of surviving individuals in a cohort $N$ follows a Gompertz relationship \citep{CalderIII:1983jd}, where
\begin{equation}
    N = N_0 {\rm exp}\left( \frac{q_0}{q_a}\Big(1 - {\rm exp}({-q_a t})\Big)\right),
\end{equation}
given that $q_0$ is the initial cohort mortality rate, and $q_a$ is the annual rate of increase in mortality, or the actuarial mortality rate.
The change in the cohort's population over time then follows
\begin{equation}
    \frac{\rm d}{\rm dt}N = -d N,
\end{equation}
such that
\begin{equation}
    d = -\frac{1}{N}\frac{\rm d}{\rm dt}N.
\end{equation}
If $t_{\rm \ell}$ is the expected lifetime of the organism, then the average rate of mortality over a lifetime $t_\ell$ is
\begin{align}
    \mu &= \frac{1}{t_{\rm \ell}}\int_0^{t_{\rm \ell}}q_0{\rm exp}(q_a t_\ell) \nonumber \\ 
    &= \frac{q_0}{q_a t_\ell}\Big({\rm exp}(q_a t_\ell) - 1\Big).
\end{align}

The cohort mortality rate $q_0$, the actuarial mortality rate $q_a$ and the expected lifetime $t_\ell$ of a mammal with mass $M_C$ all follow allometric relationships, where $q_0 = 1.88\times 10^{-8} M_C^{-0.56}$ (1/s) and $q_a = 1.45\times 10^{-7} M_C^{-0.27}$ (1/s) where $M_C$ is in grams.
Together, we obtain the allometric relationship
\begin{equation}
    \mu(M_C) =\frac{3.21\times 10^{-8} \left({\rm exp}({0.586 M_C^{0.03}})-1\right)}{M_C^{0.59}}.
\end{equation}

\subsection{Predator-prey mass relationship}
\label{sec:ppmr}

\begin{figure}[ht]
    \centering
    \includegraphics[width=1\linewidth]{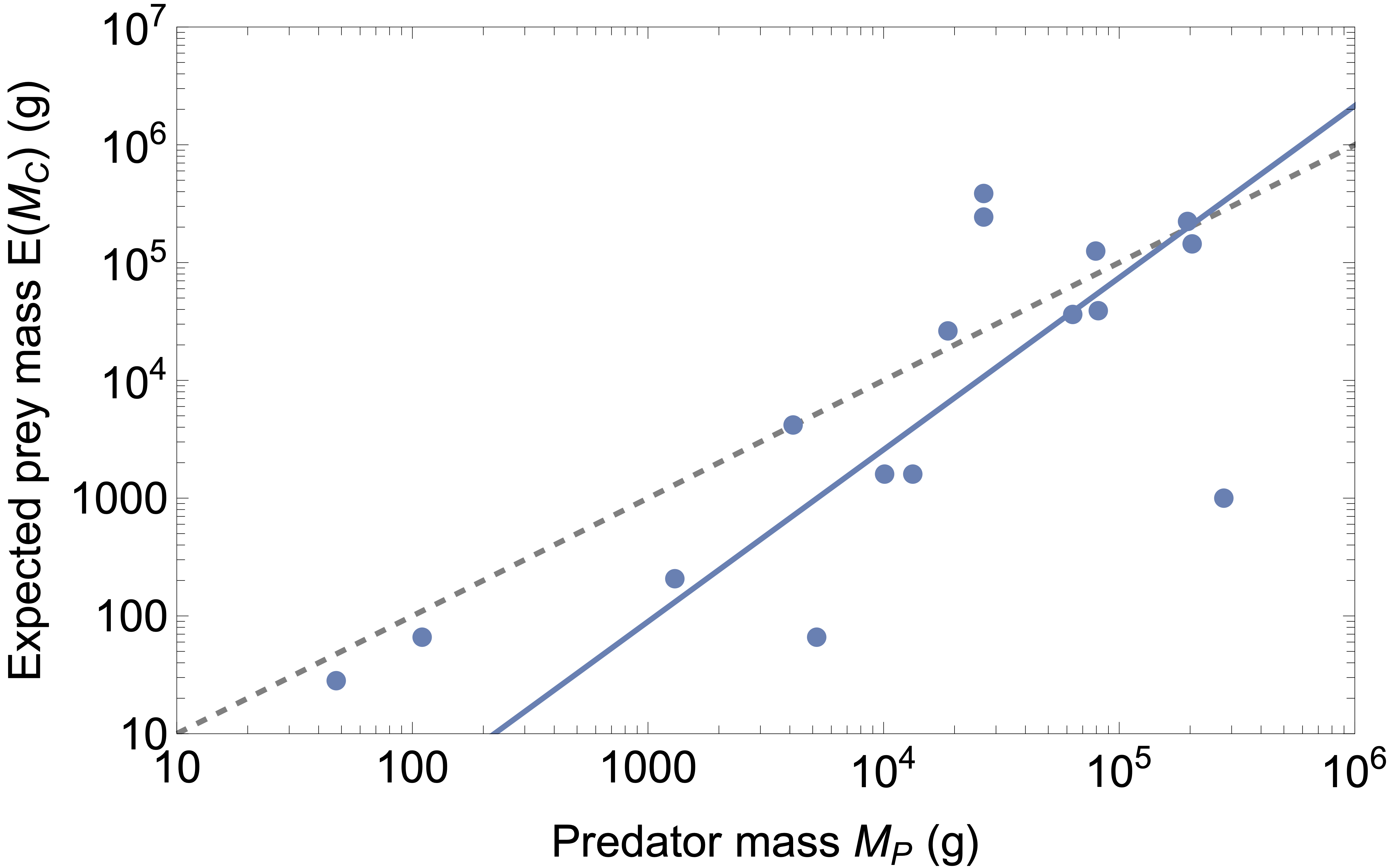}
    \caption{
    \footnotesize{
    The expected prey size for a given predator size ${\rm E}\{M_H|M_P\}$. The gray dashed line denotes the linear relationship described by Carbone et al. \citep{Carbone:1999ju}, and the blue line denotes the fitted relationship (slope: 1.46) using the data sources described in the main text.
    }
    }
    \label{fig:ppmr}
\end{figure}

To obtain the expected prey mass for a given predator mass ${\rm E}\{M_H|M_P\}$, we combined mammalian trophic relationships documented in the FoRAGE database \citep{uiterwaal2018data} with those for larger mammalian predators documented in numerous publications by Hayward et al. \citep{hayward2005lion,Hayward2006hyena,hayward2006leopard,hayward2006lycaon,hayward2006cheetah,Hayward2008}.
While previous accounts have estimated the mammalian PPMR to have a slope (in log-log space) closer to unity, the inclusion of trophic relationships for larger predators and prey, which are weighted by prey preference, suggests a slope closer to 1.46, which is the predator-prey mass relationship (PPMR) used in our analyses (Fig. \ref{fig:ppmr}).
Exclusion of grizzly bears (\emph{Ursus arctos}, which have a very low expected prey mass due to their specialization on salmon (ca. 1 kg) elevates the slope to 1.5, though we note that such minor changes to the PPMR do not impact our findings.
The expectation used throughout is set as ${\rm E}\{M_H|M_P\} = 3.73\times10^{-6} M_P^{1.46}$ with units of kg.

\subsection{Predator subsidization thresholds}
\label{sec:subsidy}

\begin{figure}[ht]
    \centering
    \includegraphics[width=1\linewidth]{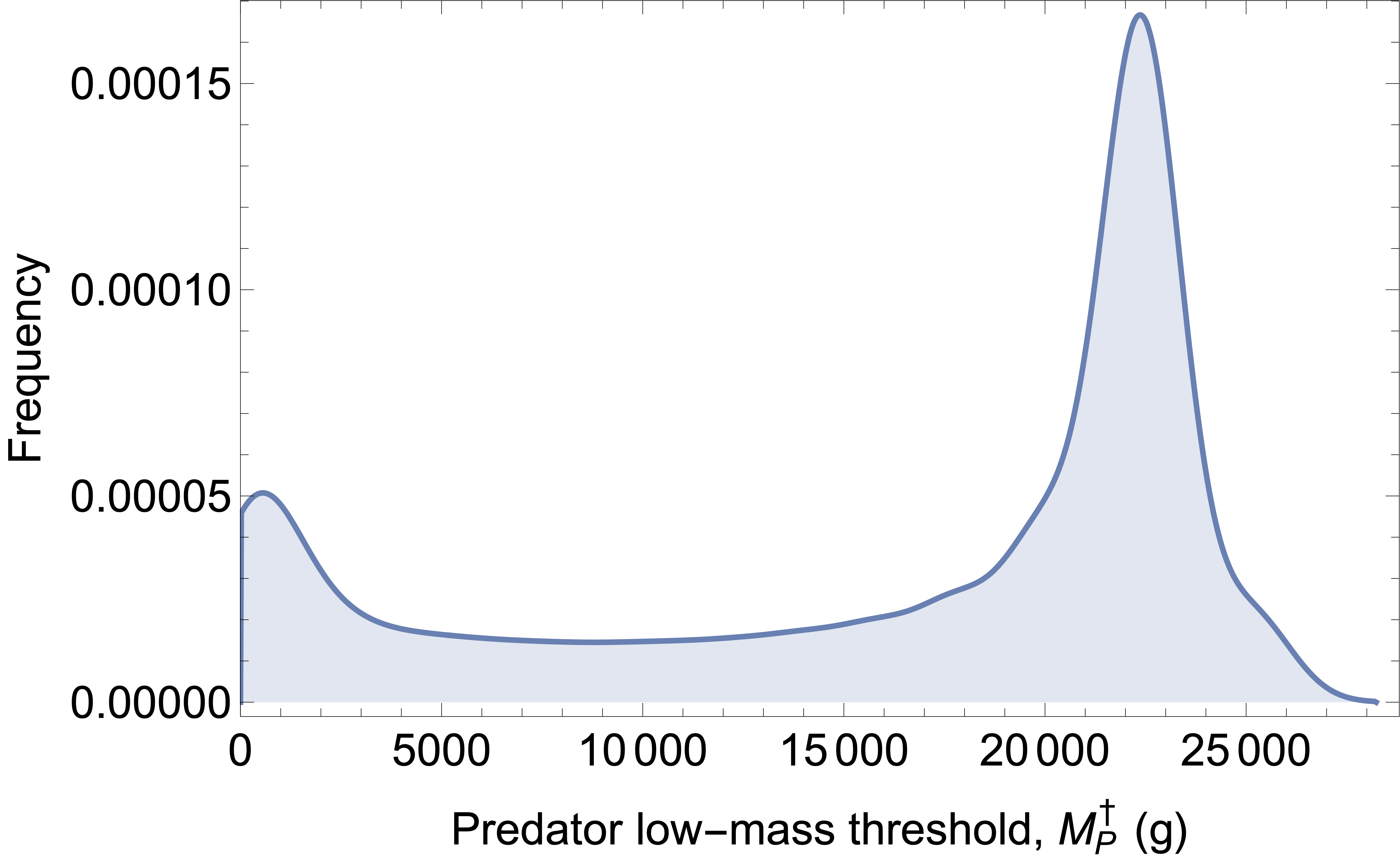}
    \caption{
        Frequency distribution of the low mass predator threshold $M_P^\dagger$ when dependence on the subsidy is low, such that $w_S < 0.1$, meaning that the predator is primarily relying on the herbivore prey.
        The average low mass threshold under these conditions is $\langle M_P^\dagger \rangle = 15$ kg, with the primary mode at $M_P^\dagger = 22$ kg.
    }
    \label{fig:predthresholdhist}
\end{figure}

The lower-mass predator threshold $M_P^\dagger$ defines the minimum predator body size that can sustain a positive population density.
Under the conditions of subsidization, we examine the frequency distribution of this lower-mass threshold when the predator is specializing on mammalian prey (such that $w_S<0.1$).
The lower-mass threshold distribution (Fig. \ref{fig:predthresholdhist}) thus defines a mammalian predator's lower body size limit when it specializes on mammalian prey.
This distribution has a peak mode at $M_P^\dagger = 22$ kg, and a long-tail towards lower $M_P^\dagger$ values, with a mean of 15 kg.
Predators with a body size lower than this threshold mass require increased subsidization to support their populations.
This cut-off aligns closely with the vertebrate-specialization threshold predicted and measured by Carbone et al. \citep{Carbone:1999ju} at ca. 21 kg.
Importantly, the prediction of the vertebrate-specialization threshold was obtained using energetic constraints based on intake rates, such that our dynamic population approach offers independent verification of this trophic constraint for mammalian predators.


\subsection{Density-dependent switching}
\label{sec:switching}

\begin{figure}[ht!]
    \centering
    \includegraphics[width=1\linewidth]{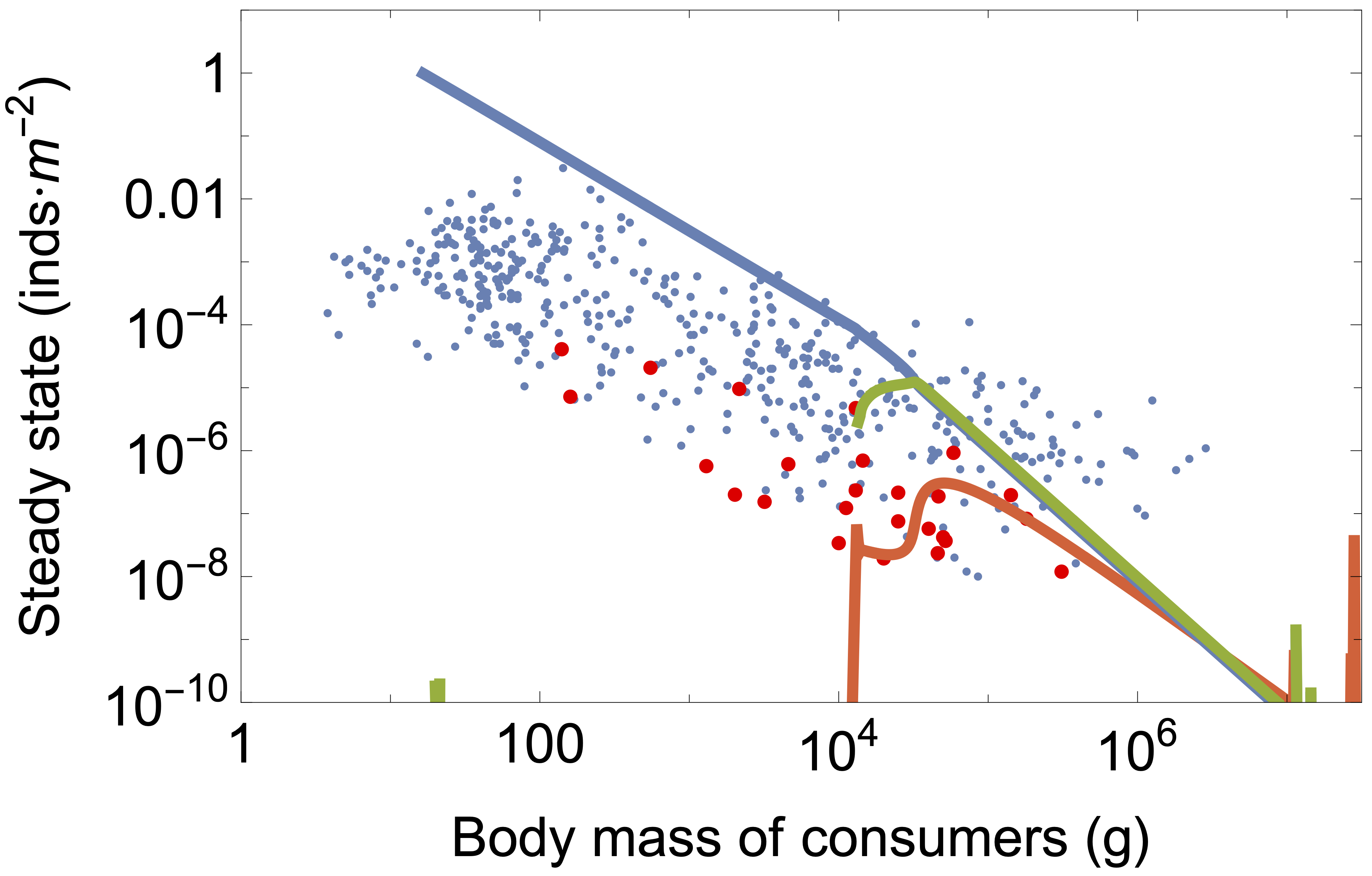}
    \caption{
        The mass-density relationship for the primary and secondary herbivore (blue and green, respectively), and the predator (red). Densities are elevated but follow a similar scaling as the tri-trophic competition model (refer to Figure 2a, main text).
    }
    \label{fig:adaptivedensities}
\end{figure}

We have assumed throughout that the proportional contribution of the primary prey to the predator ($w_1$) is constant, such that short-term fluctuations in foraging are set aside to focus on long-term, average effects on species survival and reproduction.
So while year-to-year changes in predator foraging respond to immediate changes in prey populations, it is the prey dynamics over decades and centuries shaping the average predator response, and this scale at which macroevolutionary insight is obtained.
As such, we suggest it is the averaged response of a foraging strategy, typified by the exclusion of shorter-term ecological fluctuations, that better captures the dynamics at work in shaping constraints on macroevolutionary process. 
However, we may ask to what extent our findings hold if we assume that the proportional contribution of each herbivore species changes in direct proportion to their abundances, which should not be confused with a true adaptive response \citep{valdovinos2010consequences,Kondoh2003,valdovinos2023bioenergetic}.
As such, we can allow $w_1 = \nu H_1/\sum_j H_j$ where we interpret $\nu$ to be the responsiveness to relative changes in primary prey abundance.
We observe that the inclusion of density-dependent prey-switching does not substantially change the dynamic regimes introduced by the competition model, with both predator low-mass thresholds $M_P^\dagger$ and consumer high-mass thresholds $M_H^\ddagger$ readily apparent (Fig. \ref{fig:adaptivedensities}).
Notably, density-dependent prey-switching does elevate predator population densities far above the observed mass-density relationship, as the predator is effectively feeding on a cumulatively larger prey reserve.

The size-dependent benefits of predator generalization can also be observed when $w_1$ is itself a product of the relative abundance of prey.
Because in this case, $w_1$ is no longer a fixed value, we are unable to directly distinguish a specialist from a generalist in the same way as in the main text.
Instead, we might surmise that a generalist is more responsive to changes in relative prey abundance, whereas a specialist is less responsive, such that $\nu_{\rm gen} > \nu_{\rm spec}$.
For exploratory purposes here, we will assume that the generalist predator has a prey-switching rate $10\times$ that of a specialist predator.

Dynamic predator diets typically generate extreme transient fluctuations \citep{Kondoh2003}, where the risk that a large fluctuation induces extinction can be significant.
We measure the effect that different predators have on the transient dynamics of their prey as the transient impact 
\begin{equation}
\mathcal{I} = 1 - \frac{H_{\rm low}}{H^*},
\label{eq:transientimpact}
\end{equation}
where $H^*$ is the post-transient steady state of the herbivore population density, and $H_{\rm low}$ is the lowest transient condition of the fluctuating herbivore population prior to settling to the steady state.
If the transient drop in herbivore population density is extreme, $H_{\rm low}\rightarrow0$ and $\mathcal{I} \approx 1$, meaning that the impact is maximal and will likely result in the elimination of the prey population.
If the transient drop in the herbivore population is minimal, $\mathcal{I} \approx 0$.

\begin{figure}
    \centering
    \includegraphics[width=1\linewidth]{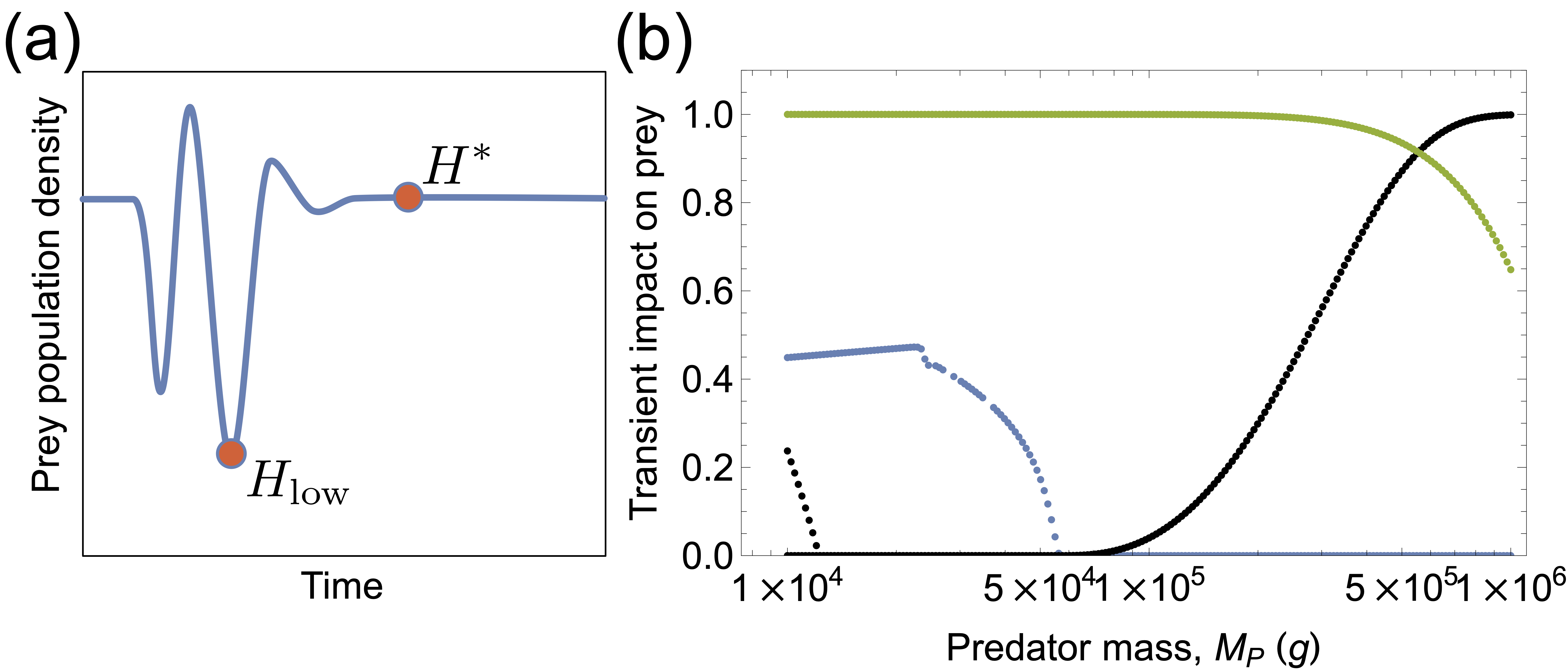}
    \caption{
        (a) The transient oscillations of an herbivore prey with respect to a generalist or specialist predator, where the lowest density is denoted as $H_{\rm low}$, and the post-transient steady state density is denoted as $H^*$.
        (b) The impact of a generalist predator versus specialist predator on the transient dynamics of the primary and secondary herbivore prey. 
        A generalist predator is defined here by one that has a high rate of prey-switching, such that $\nu_{\rm gen}>\nu_{\rm spec}$.
        The effect of the generalist predator on both herbivore prey as a function of its body size is shown in black; the effect of a specialist predator differs across prey, and is shown for the primary prey in blue and the secondary prey in green.
        Transient impact is defined in Equation \ref{eq:transientimpact}.
        As the predator body mass increases, the generalist predator tends to have a lesser impact on the transient dynamics of its prey until the change-point body size of $M_P = 549$ kg, above which the specialized predator has a lower transient impact.
    }
    \label{fig:adaptiveadvantage}
\end{figure}

We find that generalists relative to specialists produce smaller transients in their prey for predators up to the body size $M_P = 549$ kg, above which specialist predators generate a lower transient impact (Figure \ref{fig:adaptiveadvantage}).
Importantly, the change-point marking a switch from a generalist to specialist advantage is similar to that measured for extreme specialization in the non-adaptive framework presented in the main text (530 kg).
While the predator mass marking the upper limit of the generalist advantage in the main text and that based on relative transient impact are very similar, we suggest that this quantitative agreement is not inherently meaningful, as the change-point can be manipulated by changing the degree of specialization (when $w_1$ is a set value), and the relative difference of the responsiveness assumed for generalist and specialist predators (when $w_1$ is dynamic).
Instead, what is significant is the qualitative alignment that different generalist advantages sunset at extreme predator body sizes, above which specialization lowers the likelihood of a trophic-induced instability.
These results suggest that there may be multiple dynamical effects at play, serving to differentiate the impact that megapredators have on their herbivore prey relative to their smaller-bodied evolutionary cousins.


\subsection{Empirical estimates of predator dietary breadth}
\label{sec:breadth}

We calculate proxies for predator generalization in two ways for contemporary and extinct predators.
As described in the main text, for extant predators, which include Serengeti cheetah, leopards, hyaena, and lions, dietary breadth is measured directly as the ratio of the prey weight range consumed by each predator relative to the prey weight range available to all predators \citep{Sinclair2003}.
In this case, relative dietary breadth is given as $b_{\rm size} = q_{\rm pred}/q_{\rm herb}$, where $q_{\rm pred}$ is the prey mass range available for individual predator species, and $q_{\rm herb}$ is the prey mass range available to all predator species.
For predators consuming a larger range of prey body sizes, $b_{\rm size} \rightarrow 1$, implying dietary generalization.

To estimate dietary breadth for extinct species, we include 33 additional predator species from 14 separate assemblages spanning the early Pleistocene to the early Holocene \citep{palmqvist2008tracing,fox2008pleistocene,koch2004effects,desantis2021dietary,trayler2015inland,fuller2020pleistocene,fuller2014ultrafiltration,coltrain2004rancho,feranec2014understanding}, including dire wolves (\emph{Aenocyon dirus}), American lions (\emph{Panthera atrox}), saber-toothed cats (\emph{Smilodon} spp.), and the short-faced bear (\emph{Arctodus simus}), which is the largest included mammalian predator at ca. $780$ kg. 
Body size estimates for all predators were gathered from the literature \citep{anyonge2006new,christiansen2005body,dantas2022estimating,figueirido2011body,flower2016new,gazin1942late,hill2023complete,koufos2018revisiting,marciszak2022panthera,palmqvist1996prey,palmqvist2002estimating,sherani2016new,sorkin2006ecomorphology}, and are specified in detail in Supplementary Data (Zenodo citation). 

For these extinct species, dietary breadth is estimated in one of two ways.
We collected carbon and nitrogen isotope values (measured as $\delta^{13}{\rm C}$ and $\delta^{15}{\rm N}$) extracted from bone collagen (protein) or carbon isotope values only from tooth enamel (bioapatite carbonate) for both predators and their potential herbivore prey, specific to each assemblage.
Assemblages for which both carbon and nitrogen isotope values were obtained were from environments dominated by ${\rm C}_3$ vegetation, whereas those for which only carbon isotope values were obtained were from mixed ${\rm C}_3-{\rm C}_4$ environments.
We estimated relative diet breadth for predators occupying ${\rm C}_3$ environments from the relative size of isotopic convex hull area of the predator (in 2D carbon-nitrogen isotopic space), such that $b_{\rm iso} = {\rm cha}_{\rm pred}/{\rm cha}_{\rm herb}$, where ${\rm cha}_{\rm pred}$ is the convex hull area of each predator species, and ${\rm cha}_{\rm herb}$ is the convex hull area of all available herbivore species in a given assemblage.
We note that convex hull area is used in lieu of standard ellipse area (calculated from isotopic covariance), or Bayesian ellipse area (taking into account the effects of sample sizes) \citep{jackson2011comparing}, to reduce the potential for over-estimating predator generality relative to the herbivore community.
Where predator species were represented by $<3$ individuals, convex hull area could not be calculated, and the maximal $\delta^{13}{\rm C}$ range was used instead, such that $b_{\rm iso} = r_{\rm pred}/r_{\rm herb}$, where $r_{\rm pred}$ is the maximal $\delta^{13}{\rm C}$ range of each predator species, and $r_{\rm herb}$ is the $\delta^{13}{\rm C}$ isotopic range of all available herbivore species in a given assemblage.
In mixed ${\rm C}_3-{\rm C}_4$ environments, the relative diet breadth of predators is similarly estimated as $b_{\rm iso} = r_{\rm pred}/r_{\rm herb}$. 

If predators rely on the full suite of available prey (to different extents, among individuals and over time), their isotopic breadth is more likely to approximate that of the available herbivore community, such that $b_{\rm iso}\rightarrow1$, measured from either the convex hull in ($\delta^{13}{\rm C}$, $\delta^{15}{\rm N}$) space, or from the $\delta^{13}{\rm C}$ range.
On the other hand, if predator individuals specialize on a particular herbivore species or set of species, it is more likely that $b_{\rm iso} < 1$.
We note that whether stable carbon isotope ratios are obtained from bone collagen, enamel, or potentially other sources, our estimate of dietary breadth will not be impacted as long as measurements are obtained from the same biological source for herbivores and carnivores alike with respect to each assemblage.
We also note that spatio-temporal differences in mammalian isotope distributions will not bias our breadth metric as long as individuals representing a given assemblage are constrained to approximately the same space and time.
Finally, because we compare the relative isotopic areas and ranges of predator species and herbivore communities directly, without accounting for overlap, we do not need to adjust isotopic values for trophic discrimination.

In addition to the carbon isotope values for Pleistocene carnivores and their associated herbivores assembled from published data, we include five heretofore unpublished predator species from Friesenhahn Cave (Texas), dating to the the Full Glacial, including \emph{Homotherium}, \emph{Smilodon floridanus}, \emph{Ursus americanus}, and \emph{Canis latrans}.
A single specimen of \emph{Aenocyon dirus} is also included in the dataset but not the analysis, given an isotopic range cannot be calculated from a single datum.
The analytical methods used to generate isotopic data from these samples is described in Koch et al. \citep{koch2004effects} and Yann et al. \citep{yann2016dietary}.


\end{bibunit}

\end{document}